\documentclass[aps,prd,groupedaddress,graphicx,nofootinbib]{revtex4-2}
\usepackage{amsmath,amssymb,graphics,graphicx,color,epsf}
\usepackage{subfigure}
\usepackage[scanall]{psfrag}  
\usepackage{tikz}

\begin{document}

\title{The average equation of state for the oscillating inflaton field of the simplest $\alpha$-attractor E-model}

\author{Chia-Min Lin}

\affiliation{Fundamental General Education Center, National Chin-Yi University of Technology, Taichung 41170, Taiwan}



\begin{abstract}
In this work, we calculate the average equation of state for the oscillating inflaton field of the simplest $\alpha$-attractor E-model. We show that the average equation of state can be solved analytically. We discover that when $\alpha$ is small, the average equation of state of the oscillating inflaton field approaches that of a cosmological constant. This is the phenomenon of oscillating inflation.

\end{abstract}
\maketitle
\large
\baselineskip 18pt

\section{Introduction}

Cosmic inflation \cite{Starobinsky:1980te, Guth:1980zm, Linde:1981mu} is a popular theme for the very early universe.
The proposal is that an accelerated phase exists in the early universe. If it lasts long enough, many problems of the hot big bang model can be solved. 
In addition to solving problems such as the flatness problem, the horizon problem, and the problem of unwanted relics, inflation can also generate primordial density/curvature perturbations from quantum fluctuations. This explains the observed temperature fluctuation of cosmic microwave background (CMB) and provides initial conditions for subsequent structure formation. On the other hand, the observed spectral tilt of the primordial density/curvature perturbations represented by the spectral index $n_s$ can be used to constrain inflation models. It is also possible for inflation to generate potentially observable primordial gravitational waves represented by the tensor-to-scalar ratio $r$. This is another useful parameter to constrain certain inflation models.

One common way to build an inflation model is to introduce a scalar field $\phi$, called the inflaton field, with a potential $V(\phi)$. One may also consider a modified Einstein gravity and bring the model into an Einstein frame via suitable conformal transformation. The result may again be parameterized by a scalar field with a canonical kinetic term and potential. Inflation models can be roughly divided into large-field models and small-field models, where large or small indicates the comparison of the variation of the scalar field during (observable) inflation with the reduced Planck mass $M_P \simeq 2.4 \times 10^{18}\mbox{ GeV}$. One interesting feature of a large-field model is its ability to produce a large tensor-to-scalar ratio $r$. These models could be tested by experiments such as observations of cosmic microwave background (CMB). Together with the spectral index $n_s$, a model can be located on the $n_s-r$ plane.

There is a class of inflation models based on supergravity called $\alpha$-attractors \cite{Kallosh:2013yoa, Galante:2014ifa, Linde:2015uga} where $\alpha$ is a free parameter. When $\alpha$ varies, the position on the $n_s-r$ plane is running around. The value of $\alpha$ can also be taken to approach zero or infinity corresponding to different attractors on the $n_s-r$ plane. One sub-class of $\alpha$-attractors is the T-models \cite{Carrasco:2015rva, Carrasco:2015pla} with the (Einstein frame) potential of the form $f^2(\tanh(\phi/\sqrt{6\alpha}M_P))$.
Another sub-class is known as the E-models \cite{Carrasco:2015rva, Carrasco:2015pla} with the potential of the form $f^2(1-\exp(-\sqrt{2/3\alpha}\phi/M_P))$. These models could fit experimental data on the $n_s-r$ plane very well, and if $r$ is detected in the future, there will be a chance to fix the parameter $\alpha$. To make such predictions, one must choose a suitable number of e-folds $N$ corresponding to the horizon exit of the pivot wavelength of those measurements. For example, a typical value is $N=60$, used in \cite{Kallosh:2013yoa}. However, $N$ depends on the post-inflationary evolution of the inflaton field. 
It would be important to clarify the behavior of the inflaton field after inflation, in particular, during its oscillation, to determine whether $N=60$ could be used safely.
In this work, we begin with an investigation of the slow-roll parameters, and we calculate the average equation of state of the oscillating inflaton field after inflation of the simplest E-model.

\section{oscillating inflaton field}

Let us consider a homogeneous scalar field $\phi$ with a canonical kinetic term and potential $V$. The energy density $\rho$ and the pressure $p$ are given by
\begin{equation}
\rho=\frac{\dot{\phi}}{2}+V,\mbox{ and} \;\;\;  p=\frac{\dot{\phi}}{2}-V.
\end{equation}
Therefore
\begin{equation}
\dot{\phi}^2=\rho+p \equiv (1+w)\rho,
\end{equation}
where $w$ is the barotropic parameter of the equation of state. If $\phi$ is the inflaton field, it might be slow-rolling during inflation and start to enter an oscillation phase after inflation. During oscillation, $\dot{\phi}^2$ varies rapidly and (almost) periodically. Consequently, $w$ also varies quickly and oscillates. One may average $w$ over the period of the oscillation to obtain $\langle w \rangle$, which might be a constant or at least varies more slowly. The averaged continuity equation is
\begin{equation}
\frac{d\rho}{dt}=-3H(1+\langle w \rangle)\rho \equiv -3H\gamma \rho,
\label{conti}
\end{equation}
where a parameter $\gamma$ is defined as
\begin{equation}
\gamma=1+ \langle w \rangle.
\end{equation}
If $\gamma$ is a constant, Eq.~(\ref{conti}) can be integrated to give
\begin{equation}
\rho  \propto a^{-3\gamma}.
\end{equation} 
Nonrelativistic (cold) matter corresponds to $\gamma=1$, and cosmological constant corresponds to $\gamma=0$.
For an oscillating scalar field $\phi$, $\gamma$ depends on the form of the potential $V(\phi)$.
Knowledge of $\gamma$ would be important to understand the evolution of the universe before or during the inflaton decay. This is also relevant to study mechanisms such as (p)reheating or baryogenesis.

If the potential is an even function of $\phi$, we have \cite{Turner:1983he}
\begin{equation}
\gamma=2 \frac{\int^{\phi_m}_0 \left( 1- \frac{V}{V_m} \right)^{1/2}d\phi}{\int^{\phi_m}_0 \left( 1- \frac{V}{V_m} \right)^{-1/2}d\phi},
\label{g}
\end{equation}
where $\phi_m$ is the amplitude of oscillation and $V(\phi_m)=V_m$.
For the potential of the form $V \sim \phi^n$ as considered in \cite{Turner:1983he}, 
\begin{equation}
\gamma =\frac{2n}{n+2}.
\label{power}
\end{equation}
Here $n=2$ corresponds to nonrelativistic (cold) matter, and $n=0$ corresponds to a cosmological constant.
Another example where $\gamma$ can be expressed in a closed-form is for the simplest $\alpha$-attractor T-model with the potential
\begin{equation}
V=V_0 \tanh^2 \left( \frac{\phi}{F} \right).
\end{equation}
It is found by the author that \cite{Lin:2023ugk}
\begin{equation}
\gamma = \frac{2}{\cosh \left( \frac{\phi_m}{F} \right)+1}.
\end{equation}
In the following, we show that $\gamma$ can also be solved analytically for the simplest $\alpha$-attractor E-model.

\section{The simplest $\alpha$-attractor E-model and the slow-roll parameters}

\begin{figure}[t]
  \centering
\includegraphics[width=0.6\textwidth]{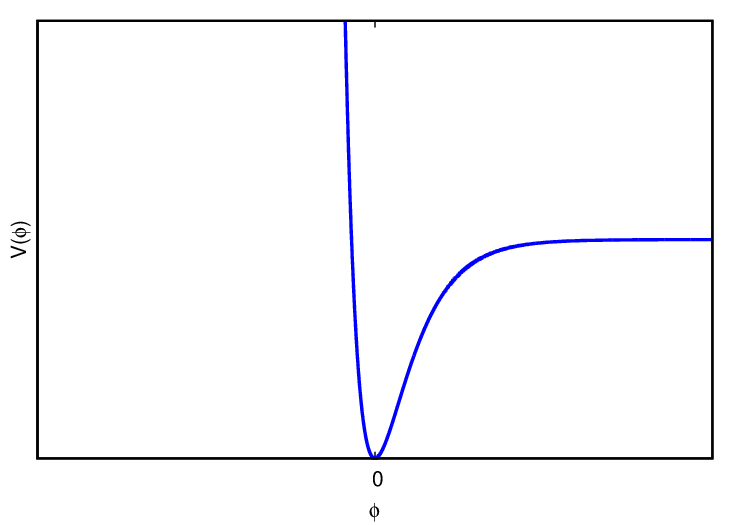}
  \caption{A typical shape of the potential $V(\phi)$ as a function of $\phi$. Inflation happens in the region $\phi>0$ where there is a plateau.}
  \label{fig3}
\end{figure}

By using a power function for $f^2(1-\exp(-\sqrt{2/3\alpha}\phi/M_P))$, the potential of the E-model is given by 
\begin{equation}
V=\Lambda^4 \left( 1-e^{-\sqrt{\frac{2}{3\alpha}}\frac{\phi}{M_P}} \right)^{2n}.
\end{equation}
This model can be built in the framework of supergravity, and the predictions as an inflation model fit the current experimental data very well.
In this work, we consider the simplest E-model, which corresponds to $n=1$, namely
\begin{equation}
V=\Lambda^4 \left( 1-e^{-\sqrt{\frac{2}{3\alpha}}\frac{\phi}{M_P}} \right)^{2} \equiv \Lambda^4 \left( 1-e^{-\lambda\phi} \right)^{2},
\label{po}
\end{equation}
where we have defined $\lambda = \sqrt{2/3\alpha M_P}$ to simplify the notation.
The model described in Eq.~(\ref{po}) is also known as the $\alpha-\beta$ model \cite{Ferrara:2013rsa, Farakos:2013cqa} which includes the Starobinsky model \cite{Starobinsky:1980te} (in the Einstein frame \cite{Whitt:1984pd}) as a special case for $\alpha=1$.
A typical shape of the potential is given in Fig.~\ref{fig3}. The potential is asymmetric under $\phi \rightarrow -\phi$. Inflation is assumed to take place in the region of $\phi>0$ where a plateau of potential exists. 

\begin{figure}[t]
  \centering
\includegraphics[width=0.6\textwidth]{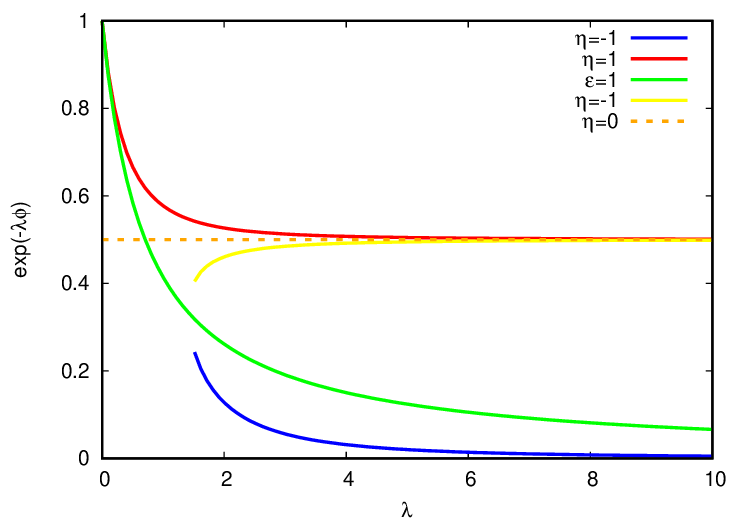}
  \caption{$e^{-\lambda\phi}$ as a function of $\lambda$ when the slow-roll fails. $\eta=-1$ occurs first when $\lambda \geq \sqrt{2}$. Otherwise, $\epsilon=1$ occurs first.}
  \label{fig2}
\end{figure}

In the following, we set the reduced Planck mass $M_P=1$.
Slow-roll inflation ends when one of the slow-roll parameters becomes unity.
From the potential given by Eq.~(\ref{po}), the slow-roll paramter $\epsilon$ is
\begin{equation}
\epsilon \equiv \frac{1}{2}\left( \frac{V^{\prime}}{V} \right)^2=\frac{2\lambda^2 e^{-2\lambda \phi}}{\left(1-e^{-\lambda \phi}\right)^2}.
\label{e}
\end{equation}
Therefore $\epsilon=1$ occurs when
\begin{equation}
e^{-\lambda \phi}=\frac{1}{1+\sqrt{2\lambda^2}}.
\end{equation}
Another slow-roll parameter $\eta$ is
\begin{equation}
\eta \equiv \frac{V^{\prime\prime}}{V}=\frac{4\lambda^2e^{-2\lambda \phi}-2\lambda^2 e^{-\lambda \phi}}{\left(1-e^{-\lambda \phi}\right)^2}.
\label{eta}
\end{equation}
Therefore $\eta=1$ occurs when
\begin{equation}
e^{-\lambda \phi}=\frac{1}{1-\lambda^2+\lambda\sqrt{\lambda^2+2}}.
\end{equation}
On the other hand, $\eta=-1$ occurs when
\begin{equation}
e^{-\lambda \phi}=\frac{\lambda^2+1 \pm \lambda\sqrt{\lambda^2-2}}{4\lambda^2 +1}.
\label{el}
\end{equation}
During inflation, the field value $\phi$ decreases. Therefore for a fixed $\lambda$, the factor $e^{-\lambda\phi}$ increases from a presumably small value. 
The question is which one comes first, $\eta=-1$, $\eta=1$, or $\epsilon=1$? We make a plot in Fig.~\ref{fig2}.
As can be seen from the figure, when $\lambda \geq \sqrt{2}$, $\eta=-1$ occurs first. For smaller $\lambda$, $\eta=-1$ never occurs and $\epsilon=1$ happens first. Note that between $\eta=-1$ (the yellow line) and $\eta=1$ (the red line), there is $\eta=0$ (the dotted orange line), which corresponds to an inflection point. From Eq.~(\ref{eta}), this happens when $e^{-\lambda\phi}=0.5$. Between two $\eta=-1$ lines (the blue line and the yellow line), we have $\eta<-1$. This implies that once $\eta=-1$ (the blue line) is achieved, the slow-roll condition fails. There is no chance to restore slow-roll conditions since $\epsilon=1$ happens before we could have returned to $\eta>-1$.

How about the region with $\phi<0$? When the inflaton field oscillates, it oscillates from a local maximum field value $\phi = \phi_{m1}>0$ (with amplitude $\phi_{m1}$) to a local minimum field value $\phi=\phi_{m2}<0$ (with amplitude $|\phi_{m2}|$). If the oscillation frequency $\omega$ is much larger than the Hubble parameter $H$, we can ignore $H$ for a single period of oscillation. From energy conservation, the relation between $\phi_{m1}$ and $\phi_{m2}$ is
\begin{equation}
V(\phi_{m1})=\Lambda^4\left( 1-e^{-\lambda \phi_{m1}} \right)^2   =V(\phi_{m2})=\Lambda^4\left( 1-e^{-\lambda \phi_{m2}} \right)^2 =V_m.
\label{lr}
\end{equation}
For the following discussion, let us define parameters $u \equiv e^{-\lambda \phi_{m1}}$ and $v \equiv e^{-\lambda \phi_{m2}}$ to make the notation simpler. Note that $u<1<v$ since $\phi_{m1}>0>\phi_{m2}$. The relation given by Eq.~(\ref{lr}) implies $u+v=2$.

As can be seen from Fig.~\ref{fig3}, the potential is steeper for $\phi<0$, and intuitively, one suspects that if the inflaton field is not slow-rolling at $\phi_{m1}$, probably it is also not slow-rolling at $\phi_{m2}$. We can be more precise here. 
Let us denote the slow-roll parameters $\epsilon = \epsilon_{+}$ when $\phi>0$ and $\epsilon = \epsilon_{-}$ when $\phi<0$. They are given by Eq.~(\ref{e}) as
\begin{equation}
\epsilon_{+}=\frac{2\lambda^2 u^2}{(1-u)^2}, 
\end{equation}
and
\begin{equation}
 \epsilon_{-}=\frac{2\lambda^2 v^2}{(1-v)^2}.
\end{equation}
Therefore $v>u$ implies $\epsilon_{-}>\epsilon_{+}$. This means when $\epsilon_{+}>1$, we must have $\epsilon_{-}>1$. Similarly, from Eq.~(\ref{eta}), the slow-roll parameters are denoted as $\eta = \eta_{+}$ when $\phi>0$ and $\eta = \eta_{-}$ when $\phi<0$. They are given by Eq.~(\ref{eta}) as
\begin{equation}
\eta_{+}=\frac{2\lambda^2 u(2u-1)}{(1-u)^2},
\end{equation}
and
\begin{equation}
\eta_{-}=\frac{2\lambda^2 v(2v-1)}{(1-v)^2}.
\end{equation}
Similarly, one can show that $\eta_{-}>\eta_{+}$. Therefore, for $\eta_{+}>1$, we have $\eta_{-}>1$. On the other hand, we can show that when $\eta_{+}<-1$, we have $\eta_{-}>1$. The conclusion is that indeed if the inflaton field is not slow-rolling at $\phi_{m1}$, it is not slow-rolling at the corresponding $\phi_{m2}$.

Let us denote the end of inflation by $\phi=\phi_e$. As can be seen from Eq.~(\ref{el}), when $\lambda \rightarrow \infty$ (which corresponds to $\alpha \rightarrow 0$), we have $\lambda \phi_e \rightarrow \infty$. If we assume the inflaton field oscillates rapidly (in the sense that $\omega \gg H$) with amplitude $\phi_m$ soon after inflation ends, $\lambda\phi_m$ can also be arbitrarily large for sufficiently small $\alpha$. This concept is important for the following section.

\section{oscillation of the inflaton field of the simplest E-model}

\begin{figure}[t]
  \centering
\includegraphics[width=0.6\textwidth]{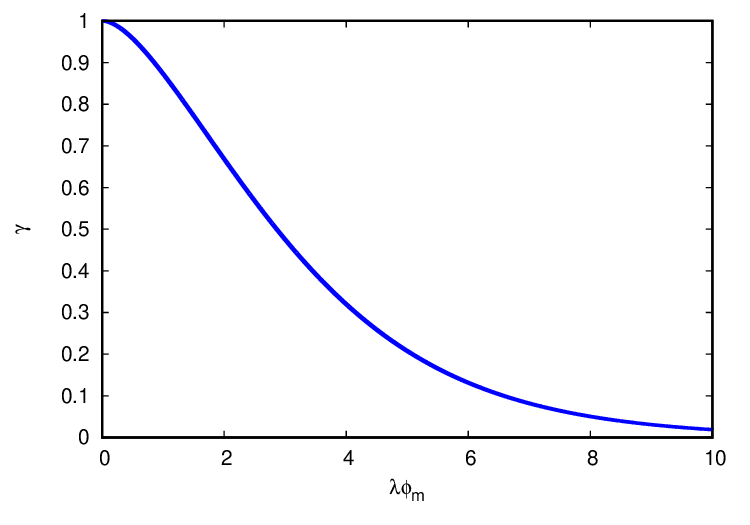}
  \caption{$\gamma$ as a function of $\lambda\phi_m$. $\gamma=1$ corresponds to the equation of state of nonrelativistic (cold) matter. $\gamma=0$ corresponds to the equation of state of the cosmological constant (oscillating inflation).}
  \label{fig1}
\end{figure}

Let us calculate $\gamma$.
The potential given by Eq.~(\ref{po}) is not an even function of $\phi$, therefore Eq.~(\ref{g}) should be modified into
\begin{equation}
\gamma=2 \frac{\int^{\phi_{m1}}_{\phi_{m2}} \left( 1- \frac{V}{V_m} \right)^{1/2}d\phi}{\int^{\phi_{m1}}_{\phi_{m2}} \left( 1- \frac{V}{V_m} \right)^{-1/2}d\phi},
\label{ga2}
\end{equation}
where $\phi_{m1}$ and $\phi_{m2}$ are given by Eq.~(\ref{lr}). For notational simplicity, let us write the integral in the numerator as
\begin{equation}
\int^{\phi_{m1}}_{\phi_{m2}} \left( 1- \frac{\Lambda^4 \left( 1-e^{-\sqrt{\frac{2}{3\alpha}}\frac{\phi}{M_P}} \right)^2}{V_m} \right)^{1/2}d\phi \equiv \int^{\phi_{m1}}_{\phi_{m2}} \left( 1-c \left(1-e^{-\lambda \phi} \right)^2 \right)^{1/2} d\phi,
\end{equation}
where
\begin{equation}
c \equiv \frac{\Lambda^4}{V_m} = \frac{1}{\left( 1-e^{-\lambda \phi_m} \right)^2}.
\label{c}
\end{equation}
Note that $c>1$.
Interestingly, the integral can be solved analytically to give
\begin{equation}
-\lambda\int^{\phi_{m1}}_{\phi_{m2}} \left( 1-c \left(1-e^{-\lambda \phi} \right)^2 \right)^{1/2} d\phi=-\pi \sqrt{c} +2\sqrt{c-1} \left( \tan^{-1} \left( \frac{1+\sqrt{c}}{\sqrt{c-1}} \right)-\tan^{-1}\left( \frac{1-\sqrt{c}}{\sqrt{c-1}} \right) \right).
\end{equation}
Note that we have multiplied $-\lambda$ to make the expression a little bit simpler.
Similarly, the integral in the denominator of Eq.~(\ref{ga2}) can be written as
\begin{equation}
\int^{\phi_{m1}}_{\phi_{m2}} \left( 1- \frac{\Lambda^4 \left( 1-e^{-\sqrt{\frac{2}{3\alpha}}\frac{\phi}{M_P}} \right)^2}{V_m} \right)^{-1/2}d\phi \equiv \int^{\phi_{m1}}_{\phi_{m2}} \left( 1-c \left(1-e^{-\lambda \phi} \right)^2 \right)^{-1/2} d\phi.
\end{equation}
Fortunately, the integral can once again be done analytically. It gives
\begin{equation}
-\lambda\int^{\phi_{m1}}_{\phi_{m2}} \left( 1-c \left(1-e^{-\lambda \phi} \right)^2 \right)^{-1/2} d\phi=\frac{2}{\sqrt{c-1}}\left( \tan^{-1} \left( \frac{1-\sqrt{c}}{\sqrt{c-1}} \right)-\tan^{-1}\left( \frac{1+\sqrt{c}}{\sqrt{c-1}} \right) \right).
\end{equation}
By substituting the results of the numerator and denominator into Eq.~(\ref{ga2}), we obtain
\begin{equation}
\gamma = 2 \frac{-\pi \sqrt{c} +2\sqrt{c-1} \left( \tan^{-1} \left( \frac{1+\sqrt{c}}{\sqrt{c-1}} \right)-\tan^{-1}\left( \frac{1-\sqrt{c}}{\sqrt{c-1}} \right) \right)}{\frac{2}{\sqrt{c-1}}\left( \tan^{-1} \left( \frac{1-\sqrt{c}}{\sqrt{c-1}} \right)-\tan^{-1}\left( \frac{1+\sqrt{c}}{\sqrt{c-1}} \right) \right)}.
\label{main}
\end{equation}
In addition, by using Eq.~(\ref{c}), we can find $\gamma$ as a function of $\lambda \phi_m$. This is plotted in Fig.~\ref{fig1}.
It can be seen from Eqs.~(\ref{c}) and (\ref{main}) that when $\lambda\phi_m \rightarrow 0$, we have $c \rightarrow \infty$ and $\gamma \rightarrow 1$. This stands to reason, since when $\lambda\phi_m \rightarrow 0$, the potential of Eq.~(\ref{po}) becomes just of a quadratic form $V \propto \phi^2$. From Eq.~(\ref{power}), we can see that $\gamma=1$ when $n=2$. On the other hand, when $\lambda\phi_m \rightarrow \infty$, we have $c \rightarrow 1$ and $\gamma \rightarrow 0$. This case is perhaps more interesting since it implies $\langle w \rangle \rightarrow -1$. The equation of state becomes that of a cosmological constant even when the inflaton field is oscillating. This phenomenon is known as oscillating inflation \cite{Damour:1997cb}. Usually, it is imagined that after slow-roll inflation and during inflaton oscillating, the scales that exit the horizon start to re-enter the horizon; namely, the comoving wavelength starts to increase. However, this is not the case if oscillating inflation happens.

\section{conclusion}
\label{con}

Our investigation of the simplest $\alpha$-attractor E-model with a potential given by Eq.~(\ref{po}) begins with a careful calculation of the slow-roll parameters. We then studied the oscillation of the inflaton field and found the corresponding $\gamma$, which determines the average equation of state. The main result is given by Eq.~(\ref{main}). Although the expression may look slightly complicated, it is nice to have an analytic solution, and the asymptotic behavior is clear. During the oscillation, $\gamma$ may vary in the range $(0,1)$. There is a similar range for the simplest $\alpha$-attractor T-model \cite{Lin:2023ugk}. 

The expression of the average equation of state $\langle w \rangle =\gamma-1$ can be applied to the study of post-inflationary evolutions of the simplest E-models for any $\alpha$. Our work provides a rare example of an exactly solvable average equation of state for a coherently oscillating scalar field. The result should be relevant to the study of (p)reheating or baryogenesis in these models.

One interesting feature of our result is that oscillating inflation could happen for a large $\lambda$ (which corresponds to a small $\alpha$). This can affect the number of e-folds $N$ used in the $n_s-r$ plane. For example, even if the number of e-folds produced during oscillating inflation is small (such as the case considered in \cite{Lin:2023ugk}), it would differ from the case of matter domination where the scales outside the horizon are entering. On the other hand, if the number of e-folds produced during oscillating inflation is not small, $N$ would need to be modified according to $\alpha$. This may affect the attractor of $\alpha \rightarrow 0$. 

\acknowledgments
This work is supported by the National Science and Technology Council (NSTC) of Taiwan under grant numbers NSTC 111-2112-M-167-002 and NSTC 112-2112-M-167-001-MY2.


\begin{thebibliography}{99}


\bibitem{Starobinsky:1980te}
A.~A.~Starobinsky,
``A New Type of Isotropic Cosmological Models Without Singularity,''
Phys. Lett. B \textbf{91}, 99-102 (1980)
doi:10.1016/0370-2693(80)90670-X

\bibitem{Guth:1980zm}
A.~H.~Guth,
``The Inflationary Universe: A Possible Solution to the Horizon and Flatness Problems,''
Phys. Rev. D \textbf{23}, 347-356 (1981)
doi:10.1103/PhysRevD.23.347

\bibitem{Linde:1981mu}
A.~D.~Linde,
``A New Inflationary Universe Scenario: A Possible Solution of the Horizon, Flatness, Homogeneity, Isotropy and Primordial Monopole Problems,''
Phys. Lett. B \textbf{108}, 389-393 (1982)
doi:10.1016/0370-2693(82)91219-9

\bibitem{Kallosh:2013yoa}
R.~Kallosh, A.~Linde and D.~Roest,
``Superconformal Inflationary $\alpha$-Attractors,''
JHEP \textbf{11}, 198 (2013)
doi:10.1007/JHEP11(2013)198
[arXiv:1311.0472 [hep-th]].

\bibitem{Galante:2014ifa}
M.~Galante, R.~Kallosh, A.~Linde and D.~Roest,
``Unity of Cosmological Inflation Attractors,''
Phys. Rev. Lett. \textbf{114}, no.14, 141302 (2015)
doi:10.1103/PhysRevLett.114.141302
[arXiv:1412.3797 [hep-th]].

\bibitem{Linde:2015uga}
A.~Linde,
``Single-field $\alpha$-attractors,''
JCAP \textbf{05}, 003 (2015)
doi:10.1088/1475-7516/2015/05/003
[arXiv:1504.00663 [hep-th]].

\bibitem{Carrasco:2015rva}
J.~J.~M.~Carrasco, R.~Kallosh and A.~Linde,
``Cosmological Attractors and Initial Conditions for Inflation,''
Phys. Rev. D \textbf{92}, no.6, 063519 (2015)
doi:10.1103/PhysRevD.92.063519
[arXiv:1506.00936 [hep-th]].

\bibitem{Carrasco:2015pla}
J.~J.~M.~Carrasco, R.~Kallosh and A.~Linde,
``$\alpha $-Attractors: Planck, LHC and Dark Energy,''
JHEP \textbf{10}, 147 (2015)
doi:10.1007/JHEP10(2015)147
[arXiv:1506.01708 [hep-th]].


\bibitem{Turner:1983he}
M.~S.~Turner,
``Coherent Scalar Field Oscillations in an Expanding Universe,''
Phys. Rev. D \textbf{28}, 1243 (1983)
doi:10.1103/PhysRevD.28.1243

\bibitem{Lin:2023ugk}
C.~M.~Lin,
``On the oscillations of the inflaton field of the simplest $\alpha$-attractor T-model,''
[arXiv:2303.13008 [hep-ph]].

\bibitem{Ferrara:2013rsa}
S.~Ferrara, R.~Kallosh, A.~Linde and M.~Porrati,
``Minimal Supergravity Models of Inflation,''
Phys. Rev. D \textbf{88}, no.8, 085038 (2013)
doi:10.1103/PhysRevD.88.085038
[arXiv:1307.7696 [hep-th]].

\bibitem{Farakos:2013cqa}
F.~Farakos, A.~Kehagias and A.~Riotto,
``On the Starobinsky Model of Inflation from Supergravity,''
Nucl. Phys. B \textbf{876}, 187-200 (2013)
doi:10.1016/j.nuclphysb.2013.08.005
[arXiv:1307.1137 [hep-th]].



\bibitem{Whitt:1984pd}
B.~Whitt,
``Fourth Order Gravity as General Relativity Plus Matter,''
Phys. Lett. B \textbf{145}, 176-178 (1984)
doi:10.1016/0370-2693(84)90332-0

\bibitem{Damour:1997cb}
T.~Damour and V.~F.~Mukhanov,
``Inflation without slow roll,''
Phys. Rev. Lett. \textbf{80}, 3440-3443 (1998)
doi:10.1103/PhysRevLett.80.3440
[arXiv:gr-qc/9712061 [gr-qc]].

\end{thebibliography}
\end{document}